\documentclass[twocolumn,prl,showpacs,color,superscriptaddress,epsfig,floatfix,amsmath,amssymb,amsfonts]{revtex4}
\usepackage{graphicx}
\usepackage{color}
\usepackage{multirow}

\begin{document}

\title{Modified kagom\'{e} physics in the natural spin-1/2 kagom\'{e} lattice
systems --- kapellasite Cu$_3$Zn(OH)$_6$Cl$_2$ and haydeeite
Cu$_3$Mg(OH)$_6$Cl$_2$}

\author{O. \surname{Janson}}
\affiliation{Max-Planck-Institut f\"{u}r Chemische Physik fester
Stoffe, D-01187 Dresden, Germany}

\author{J. \surname{Richter}}
\affiliation{Institut f\"{u}r Theoretische Physik, Universit\"{a}t
Magdeburg, D-39016 Magdeburg, Germany}

\author{H. \surname{Rosner}}
\email{rosner@cpfs.mpg.de}
\affiliation{Max-Planck Institut f\"{u}r Chemische Physik fester
Stoffe, D-01187 Dresden, Germany}

\date{\today}

\begin{abstract}
The recently discovered natural minerals Cu$_3$Zn(OH)$_6$Cl$_2$ and
Cu$_3$Mg(OH)$_6$Cl$_2$ are spin 1/2 systems with an ideal kagom\'{e}
geometry. Based on electronic structure calculations, we develop a
realistic model which includes couplings across the kagom\'{e}
hexagons beyond the original kagom\'{e} model that are intrinsic in
real kagom\'{e} materials. Exact diagonalization studies for the
derived model reveal a strong impact of these couplings on the
magnetic ground state. Our predictions could be compared to and
supplied with neutron scattering, thermodynamic and NMR data.

\end{abstract}


\maketitle

For decades, low-dimensional spin systems attract broad interest due
to their intriguing, unusual ground states (GS) such as helically
ordered, spin Peierls, spin-liquid or resonating valence-bond
GS's~\cite{moessner01,diep04,scholl04,phys_today06}.  These unusual
GS's are typically driven by competing interactions or
geometric frustration.  Two-dimensional (2D) quantum spin systems are
of particular interest because the competition between quantum
fluctuations and interactions seems to be well balanced, and fine
tuning of this competition may lead to zero-temperature transitions
between semi-classical and quantum phases~\cite{sachdev04}. There are
several examples for strongly frustrated 2D quantum spin materials,
e.g. PbVO$_3$~\cite{PbVO3} or SrCu$_2$(BO$_3$)$_2$~\cite{kageyama99},
which can be well described by a frustrated spin-1/2 Heisenberg
model. Such 2D quantum magnets are at present most suitable objects
for the comparison between theory and experiment.

A simple but very challenging realization of a geometrically
frustrated quantum magnet is the spin-1/2 Heisenberg antiferromagnet
(HAFM) on a kagom\'{e} lattice.  The kagom\'{e} HAFM attracts much
interest due to its unusual classical and quantum GS's and
low-temperature thermodynamics, see e.g.
Refs.~\onlinecite{chalker92,huse92,reimers93,%
  waldtmann98,lhuillier01,Rigol_PRL_07}, and also due to potential
applications of a possible quantum spin-liquid
state~\cite{quantum_comp,Ying}.
The recent discovery of a natural spin-1/2 kagom\'{e} compound
Cu$_3$Zn(OH)$_6$Cl$_2$ (mineral
herbertsmithite~\cite{herbertsmithite}) and a subsequent synthesis of
good-quality samples~\cite{herbertsmithite_synth} have spurred both
experimental ~\cite{Mendels_PRL_07,Helton_PRL_07} and theoretical
~\cite{waldtmann98,lhuillier01,Rigol_PRL_07} investigations of this
frustrated magnetic system. The experimental results were quite
unexpected: Curie-Weiss behavior with a rather large $\Theta$, an
upturn in magnetic susceptibility at 75~K and no spin gap down to 100
mK are far from being consistently described by theory. The main
obstacle for theoretical studies is the structural \mbox{Cu---Zn}
disorder within this compound \cite{disorder}, which hampers the
kagom\'e physics, but encourages the search for new materials. A very
recent discovery of two isostructural spin-1/2 kagom\'{e}
systems --- the minerals kapellasite Cu$_3$Zn(OH)$_6$Cl$_2$
(\cite{kapellasite}, a metastable polymorph of herbertsmithite) and
haydeeite Cu$_3$Mg(OH)$_6$Cl$_2$~\cite{haydeeite}, widens the range of
possible investigations. These systems are of great potential interest
because (i) no cations are located between the planes, thus less
coupling between kagom\'{e} layers is expected though the interlayer
distance is reduced by about 1 \r{A} (ii) the presence of two
isostructural compounds should allow a systematic study of additional
exchange couplings beyond the original kagom\'{e} model.  We have
performed a theoretical electronic structure study within density
functional theory (DFT) and estimated the exchange parameters of a
corresponding Heisenberg model. For this spin model we have calculated
the classical GS and for a finite lattice of $N=36$ sites the quantum
spin-1/2 GS.

The DFT calculations were performed using a full-potential
nonorthogonal local-orbital scheme (FPLO version 6.00-24)~\cite{fplo}
within the local density approximation (LDA).  The Perdew and Wang
parameterization of the exchange-correlation potential was chosen for
the scalar relativistic calculations~\cite{perdew_wang}. The default basis
set was used. The strong on-site correlations of the Cu $d$-electrons
were taken into account using the LSDA+\textsl{U} method~\cite{lsdu}.
Well converged \textsl{k}-meshes of 124 points for the conventional cell and 75
points for the supercell in the irreducible wedge were used.

\begin{figure}
\includegraphics[width=5cm]{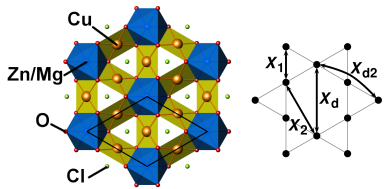}
\caption{\label{str} (Color online) The hexagonal crystal structure of
kapellasite and haydeeite: CuO$_4$ plaquettes form a buckled kagom\'{e}
layer bridged by ZnO$_6$/MgO$_6$ octahedra. The
inter-layer space is filled with Cl and H (not shown) atoms.}
\end{figure}

\begin{figure}
\includegraphics[angle=270,width=7.5cm]{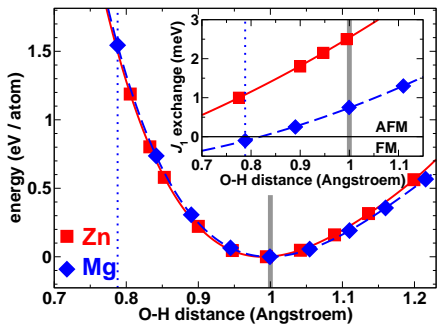}
\caption{\label{OH} (Color online) Total energy per H atom given by
LDA calculations for kapellasite (Zn) and haydeeite (Mg). The zero
energy corresponds to the calculated equilibrium distance and is marked with a
gray line.  The experimental value of the O---H distance in haydeeite is shown
with a dotted line. Inset: exchange \textsl{J}$_1$ from the supercell
LSDA+\textsl{U} calculations as a function of the O---H distance.} 
\end{figure}

The hexagonal crystal structure of both minerals consists of layers
(Fig.~\ref{str}) perpendicular to the $c$ direction. These layers are
built by a kagom\'{e} lattice of corner-sharing CuO$_4$ plaquettes,
which are tilted with respect to this plane, and ZnO$_6$ (kapellasite)
or MgO$_6$ (haydeeite) octahedra bridging the ``ring'' of six CuO$_4$
plaquettes. 
The Cu---O---Cu angle between two neighboring plaquettes is close to
105$^{\circ}$, providing considerable ferromagnetic (FM) contributions
to the exchange due to the vicinity to 90$^{\circ}$.  The kagom\'{e}
layers are separated by Cl atoms, which are bonded to H atoms that
stick out of the layers.  The experimentally defined H position for
haydeeite~\cite{mg_structure} yields the unusually short O---H distance of 0.78
\r{A}, the H position in kapellasite has not been reported.  To account for
this structural peculiarity, the H position was relaxed with respect to the
total energy, we show below that it has a dramatic impact on the exchange.
Throughout the paper, we use the optimized H position~\cite{foot1}
 yielding $\sim$1.0 \r{A}
for the O---H distance (Fig.~\ref{OH}).

\begin{figure}
\includegraphics[angle=270,width=7.5cm]{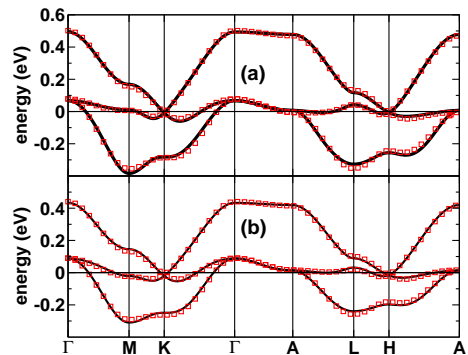}
\caption{\label{bs} 
(Color online) TB model (squares) fitted to the band structures (solid lines) of
kapellasite (a) and haydeeite (b).}
\end{figure}

Our LDA calculations yield a valence band with a total width of 6---7~eV for
both compounds with three bands crossing the Fermi level  $\varepsilon_{\rm{F}}$
according to the three Cu atoms per unit cell (Fig.~\ref{bs}). The valence
bands of haydeeite and kapellasite have two pronounced differences: (1) the
rather localized \textsl{d}-states of Zn (between $-$6.5 and $-$4~eV)
contribute to the valence band of kapellasite while Mg states have negligible
contribution for haydeeite and (2) the width of the separated band complex at
$\varepsilon_{\rm{F}}$ is slightly different. Nevertheless, the same model can be
applied for the description of low energy excitations.

The band structure (Fig.~\ref{bs}) reveals that the dispersion
perpendicular to the kagom\'{e} planes (along $\Gamma$---A) is very
small, pointing to a pronounced 2D character of the systems in
accordance with our expectations. The presence of states at Fermi
level yields a metallic GS, contrary to the insulating behavior
typical for undoped cuprates~\cite{foot2}. This discrepancy originates
from the strong on-site correlations of the Cu 3\textsl{d} electrons,
insufficiently described by LDA, and can be accounted for by adding
the missing Coulomb repulsion in a model Hamiltonian or in the
LSDA+\textsl{U} approximation. Though LDA fails to describe the
correlations correctly, it is known to provide reliable values of
transfer integrals \cite{foot3}
 which
can be used for a model analysis of the magnetic excitations.
To define the relevant orbitals for the low energy excitations we
analyzed the density of states (DOS) by applying local coordinate
systems for all orientations of CuO$_4$ plaquettes and calculated the
local DOS and band weights.  The analysis revealed that the bands at
Fermi level belong to local Cu 3\textsl{d}$_{x^2-y^2}$ and O
2\textsl{p}$_{\sigma}$ orbitals, i.e. the standard cuprate
scenario with a half-filled antibonding \textsl{dp}$\sigma^{*}$ band is
realized. Therefore, an effective one-band model, already applied
for similar materials~\cite{Sr2CuPO42,Bi2CuO4}, is appropriate to
describe the magnetic excitations in these systems.

Three \textsl{dp}$\sigma^{*}$ bands per unit cell lead to a 3$\times$3
matrix representing the tight-binding (TB) Hamiltonian. The number of
transfer integrals, included into the TB model was picked to get a
good fit of the LDA bands which could not be considerably improved by
inclusion of further parameters. The fits shown in Fig.~\ref{bs} were
achieved using ten transfer integrals, though only four of them
(Fig.~\ref{str}) were larger than 10~meV. To check the stability of
the leading terms, we subsequently decreased the number of parameters
in our model. Basing on these results, we estimate less than 10\%
uncertainty in our values for the leading four transfer integrals
depending on the chosen TB Hamiltonian. Thus, we can restrict
ourselves to analysis of the leading terms. In order to estimate
the antiferromagnetic (AF) exchange, the transfer integrals were
mapped to an extended Hubbard model and subsequently to a Heisenberg
model with
\textsl{J}$^{AF}_i=4t^2_i/U_{eff}$~\cite{foot5}.

The leading AF exchange in both systems is the nearest-neighbor (NN)
exchange \textsl{J}$^{AF}_1$, the second largest is the exchange
along diagonals of a kagom\'{e} lattice, \textsl{J}$^{AF}_d$ (see
Fig.~\ref{str}). The relevance of the latter exchange is rather
unexpected: while the pure kagom\'e model includes \textsl{J}$_1$ only,
its modifications usually contain only the second neighbor
\textsl{J}$_2$ \cite{j1j2}. In our case, we find \textsl{J}$^{AF}_{2}$ 
(and also \textsl{J}$^{AF}_{d2}$) smaller than 0.5~meV for
both systems, thus these terms can be neglected in the following discussion. The
inter-plane coupling is much smaller than 0.1~meV, showing that the
systems are almost perfect 2D magnets. Therefore, a
\textsl{J}$_1$-\textsl{J}$_d$ model should be appropriate to describe
the magnetism in good approximation.

The values of the total exchange were obtained using total energy
calculations for supercells with different spin arrangements, where
the difference in the values of total energy originates only from spin
degrees of freedom. To obtain the leading exchange integrals, we used
a doubled cell with six Cu atoms.  The supercell calculations were
performed using the LSDA+\textsl{U} method treating the
correlation on a mean field level. This approach is necessary due to
the Cu---O---Cu bond angle $\approx105^{\circ}$ which leads to
sizable FM contribution according to Goodenough-Kanamori-Anderson
rules.
\begin{figure}
\includegraphics[angle=270,width=6.5cm]{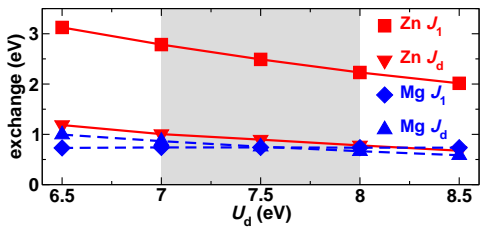}
\caption{\label{lsdu} 
(Color online) Exchange integrals of kapellasite and haydeeite as a
function of Coulomb repulsion \textsl{U}.
}
\end{figure}

\begin{table}
\begin{ruledtabular}
	\begin{tabular}{c | r r r r | r r r r}
		& \multicolumn{4}{c}{kapellasite} & \multicolumn{4}{c}{haydeeite} \\
		path & \textsl{t} & \textsl{J}$^{AF}$ & \textsl{J}$^{FM}$ & \textsl{J} & \textsl{t} & \textsl{J}$^{AF}$ & \textsl{J}$^{FM}$ & \textsl{J} \\ \hline
		 \textsl{X}$_1$ & 87 & 7.5 &  $-$5.0 & 2.5 & 73 & 5.3 & $-$4.5 & 0.8\\
		 \textsl{X}$_2$ & $-$10 & 0.1 & $\sim$0 & $<$0.1 & $-$9 & 0.1 & $\sim$0  & $<$0.1 \\
		 \textsl{X}$_{d}$ & 49 & 2.4 & $-$1.5 & 0.9 & 42 & 1.8 & $-$1.0 & 0.8 \\
		 \textsl{X}$_{d2}$ &20 & 0.4 & $-$0.4 & $<$0.1 & 22 & 0.5 & $-$0.5 & $<$0.1\\
	\end{tabular}
\caption{\label{table}Transfer and exchange integrals of kapellasite
and haydeeite. All values are given in meV. The values of transfer
integrals are taken from the TB model. The AF exchange is calculated
via subsequent mapping the transfer integrals to the extended Hubbard
and Heisenberg models. The total exchange is taken from
LSDA+\textsl{U} total energy calculations of supercells.
\textsl{J}$^{FM}$ is the difference between \textsl{J} and
\textsl{J}$^{AF}$.}
\end{ruledtabular}
\end{table}

The results are given in Table~\ref{table} (\textsl{J}$^{FM}$ was
evaluated as the difference between the total \textsl{J} and
\textsl{J}$^{AF}=4t^2/$\textsl{U}$_{eff}$).  The FM contributions
significantly modify the size of the relevant exchange integrals, but
preserve their AF nature. Here, we introduce the ratio
$\alpha\equiv$\textsl{J}$_{d}$/\textsl{J}$_1$, which is zero in the
simple kagom\'{e} model and runs to infinity in case of decoupled
chains. Certainly, $\alpha$ may depend on external parameters like the
H position and the \textsl{U} values. The change in O---H distance
drastically affects the \textsl{J}$_1$ exchange (Fig.~\ref{OH}, inset)
in both compounds, and especially in haydeeite, where it becomes FM
when the O---H bond is shorter than 0.8 \r{A}. Thus, further quantitative analysis
is based on the empirical fact that total energy calculations provide
in general rather precise atomic positions. An
accurate experimental determination of the H position is highly
desirable for an improvement. The influence of the Coulomb repulsion \textsl{U} on
$\alpha$ is much weaker and there are no drastic changes in verified
region (Fig.~\ref{lsdu}): $\alpha$ is very close to 0.36 for
kapellasite and stays in the vicinity of unity for haydeeite for the
whole range of $U$ studied. While GS's for $\alpha=0$ and
$\alpha=\infty$ are relatively clear, the region in between is not
studied. Therefore, we have performed exact diagonalization studies in
order to clarify the influence $\alpha$ on the GS.

It is well known that the classical GS of the pure kagom\'{e} HAFM ($\alpha=0$)
is highly degenerate \cite{huse92,chalker92,reimers93}.  The additional
diagonal bond \textsl{J}$_d$ reduces this degeneracy drastically and selects
non-coplanar GS's with twelve magnetic sublattices \cite{foot6}
among the huge number
of classical kagom\'{e} GS's.  These classical GS's of the
\textsl{J}$_1$-\textsl{J}$_d$ model are characterized by a perfect antiparallel
(N\'{e}el) spin alignment along the chains formed by diagonal bonds
\textsl{J}$_d$ and by a $120^{\circ}$ spin arrangement on each triangle formed
by NN bonds $J_1$. As a result, every two spin-sublattices are N\'{e}el-like
antiparallel to each other and these two sublattices are
perpendicular to one other group of two N\'{e}el-like sublattices.

\begin{figure}
\includegraphics[angle=270,width=7.8cm]{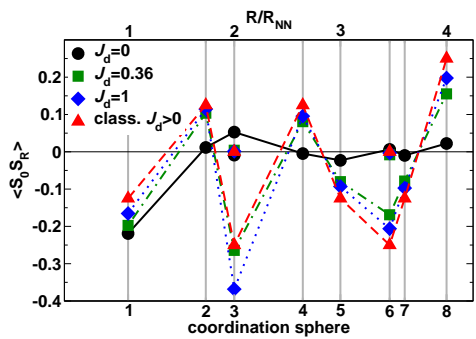}
\caption{\label{sisj} (Color online) GS spin-spin correlation $\langle
{\bf S}_0{\bf S}_{\bf R}\rangle $ versus separation R$=|{\bf R}|$ for
the \textsl{J}$_1$-\textsl{J}$_d$ HAFM on the kagom\'{e} lattice. The
results for the quantum $S=1/2$ model are calculated for a finite
lattice of $N=36$ sites. The lines are guides for the eyes connecting
data points. Note that for $\rm{R}=2$ and $\sqrt{12}$ two
non-equivalent spin-spin separations exist and the lines connect
the points representing the stronger correlation.}
\end{figure}

For the quantum model the GS and low-lying excitations have been
calculated by Lanczos diagonalization for the finite lattice of $N=36$
considered previously in the literature for the pure kagom\'{e} HAFM.
Note that this finite lattice fits to the magnetic structure of the
classical GS. The calculated spin correlations $\langle {\bf
  S}_0{\bf S}_{\bf R}\rangle $ for the classical GS as well as for the
quantum GS for $\alpha=0.36$ and $\alpha=1.0$ are shown in Fig.~\ref{sisj}.
For comparison we also show $\langle {\bf S}_0{\bf S}_{\bf R}\rangle $
for the pure kagom\'{e} system, i.e. \textsl{J}$_{d}=0$.  Obviously,
the quantum GS spin correlation is drastically changed by
\textsl{J}$_d$. While for \textsl{J}$_d=0$ the decay of the spin
correlation function is extremely rapid, we find a well pronounced
short-range order for $\alpha=0.36$ and $\alpha=1.0$ that corresponds to
the classical magnetic structure. This leads to the conclusion that
even in the quantum model the GS has a non-coplanar magnetic
structure giving rise to enhanced chiral correlations. Moreover, it is
obvious from Fig.~\ref{sisj} that the magnetic correlations along the
chains built by $J_d$ bonds (
$R/R_{NN}=2$ and $4$) are strongest, indicating that the low-energy
excitations might be $S=1/2$ spinons causing an effectively one-dimensional
low-temperature physics similar to other 2D models, e.g.
the crossed-chain model~\cite{Starykh_PRL_02} and the
anisotropic triangular lattice~\cite{Coldea_PRL_01}. However, this issue needs
further investigation.

Finally we mention another important difference from the pure kagom\'{e}
system which is relevant for the low-temperature thermodynamics.  For
$\alpha=0$ the singlet-triplet gap (spingap) is filled by 210
non-magnetic excitations~\cite{waldtmann98,lhuillier01} leading to
different low-temperature behavior of the specific heat $C$
(power-law in $T$) and the susceptibility $\chi$ (exponential
decay). By contrast we find that for $\alpha=1$ ($\alpha=0.36$) there
are no (only a few) singlets within the spingap.  Therefore we do not
expect any basic difference in the low-$T$ behavior of $C$ and $\chi$.

To summarize, we have performed electronic structure calculations for
two new spin-1/2 kagom{\'{e}} lattice compounds --- kapellasite and
haydeeite. Both compounds are 2D magnets, with two relevant AF
exchanges: NN exchange \textsl{J}$_1$ and the exchange along
``diagonals'' of a kagom{\'{e}} lattice \textsl{J}$_d$.  We find
$\alpha\equiv$\textsl{J}$_{d}$/\textsl{J}$_1\approx0.36$ for
kapellasite and $\alpha\approx1$ for haydeeite.  The exchanges and
thus $\alpha$ values are strongly dependent on the H position for
which the experimental value is unlikely with respect to the total
energy and should be reinvestigated. The presence of significant
\textsl{J}$_d$ interaction leads to (i) non-coplanar magnetic order
with 12 sublattices on the classical level which at least on a
short-range scale, shows similarities to in the quantum model, and
(ii) to the shift of the low-lying singlets out of the spingap. We
especially emphasize the crucial importance of $J_d$ for all real
materials with kagom{\'{e}} geometry, which needs careful
consideration in order to obtain the physically relevant model for the
GS and the low lying excitations.  This work is a starting point for
study of these promising model compounds.
  
Our predictions could be challenged and extended by low temperature
experiments: neutron scattering and $\mu$SR to probe the spin-spin
correlation function, thermodynamic measurements ($C$ and $\chi$, see
above) to check for a possible spin gap, including pressure studies
to modify $\alpha$ via a change of the O--H distance.

The work was supported by Emmy Noether program of DFG. We acknowledge
fruitful discussions with S.-L.~Drechsler and J.~Schulenburg.

\begin{thebibliography}{37}
\bibitem[{foo({\natexlab{a}})}]{foot1}
\bibinfo{note}{The crystal structures were determined by x-ray diffraction;
  this method usually does not allow to define the H position with good
  precision.}
\bibitem[{foo({\natexlab{e}})}]{foot5}
\bibinfo{note}{\textsl{U}$_{eff}$ = 4~eV was used. Thus, the mapping is well
  justified in the strongly correlated limit $U_{eff}{\gg}t_i$ at
  half-filling.}
\bibitem[{foo({\natexlab{f}})}]{foot6}
\bibinfo{note}{The GS's are similar to the GS shown in Fig.~2 in
  Ref.~\onlinecite{j1j2}.}
\end{thebibliography}

\end{document}